\documentclass[aps,pre,twocolumn,amsmath,amssymb,nofootinbib,superscriptaddress,showpacs,longbibliography]{revtex4-1}

\usepackage{graphicx}
\usepackage{natbib}
\usepackage{latexsym}
\usepackage{amsmath, amssymb, amsthm}
\usepackage{bbold}
\usepackage{mathtools}
\usepackage{xcolor}
\usepackage{comment}
\usepackage{wasysym}

\usepackage[bottom]{footmisc}

\usepackage{microtype}
\usepackage{epigraph}
\usepackage{lipsum}

\usepackage{float}

\usepackage[normalem]{ulem}

\usepackage{appendix}

\usepackage{hyperref}

\begin{document}

\title{Extended-range percolation in complex networks}

\author{Lorenzo Cirigliano}
\affiliation{Dipartimento di Fisica Universit\`a ``Sapienza”, P.le
  A. Moro, 2, I-00185 Rome, Italy}
  
\affiliation{Centro Ricerche Enrico Fermi, Piazza del Viminale, 1,
  I-00184 Rome, Italy}

\author{Claudio Castellano}
\affiliation{Istituto dei Sistemi Complessi (ISC-CNR), Via dei Taurini
  19, I-00185 Rome, Italy}

\affiliation{Centro Ricerche Enrico Fermi, Piazza del Viminale, 1,
  I-00184 Rome, Italy}

\author{G\'abor Tim\'ar}
\affiliation{Departamento de F\'\i sica da Universidade de Aveiro \& I3N, Campus Universit\'ario de Santiago, 3810-193 Aveiro, Portugal}

\date{\today}

\begin{abstract}
Classical percolation theory underlies many processes of
information transfer along the links of a network.
In these standard situations, the requirement for two nodes to be
able to communicate is the
presence of at least one uninterrupted path of nodes between them.
In a variety of more recent data transmission protocols,
such as the communication of noisy data via error-correcting repeaters,
both in classical and quantum networks,
the requirement of an uninterrupted path is too strict:
two nodes may be able to
communicate even if all paths between them have interruptions/gaps
consisting of nodes that may corrupt the message.
In such a case a different approach is needed.
We develop the theoretical framework
for extended-range percolation in networks, describing the fundamental
connectivity properties relevant to such models of information
transfer. We obtain exact results, for any range $R$,
for infinite random uncorrelated
networks and we provide a message-passing formulation that works well
in sparse real-world networks. The interplay of the extended range and
heterogeneity leads to novel critical behavior in scale-free networks.
\end{abstract}


\maketitle

\section{Introduction}
\label{sec1}

Percolation theory investigates how connectivity structures in a
network change if some nodes or links are removed/deactivated
\cite{stauffer2018introduction}. Standard (classical) percolation uses
a definition of connectedness whereby two active nodes belong to the same
connected component if there is at least one uninterrupted path of
active nodes between them \cite{newman2010networks, Li2021}. This
model setup provides the basis for the mathematical description of a
wide range of physical processes, such as flow through porous media
\cite{moon1995critical}, forest fires
\cite{henley1993statics}, epidemic spreading 
\cite{moore2000epidemics, Newman2002, pastor2015epidemic} and
various types of transport phenomena \cite{barrat2008dynamical, germano2006traffic,
  pei2013spreading, zhang2016dynamics}. A more recent, well-studied
class of generalizations of standard percolation use connectivity
definitions that involve multiple uninterrupted paths between nodes,
such as in $k$-core \cite{dorogovtsev2006k} or bootstrap percolation
\cite{baxter2010bootstrap} and in the mutual
connectivity rule of multiplex networks
\cite{buldyrev2010catastrophic}. These models, particularly popular in
recent years, are suitable for studying phenomena such as complex
contagion \cite{centola2007complex} or multilayer
spreading processes \cite{de2018fundamentals}.

In a variety of information spreading processes, however, connectivity
definitions based solely on uninterrupted paths are not
appropriate. In the case of noisy data transmission, for example,
signals may deteriorate while passing through the nodes of a network,
and the presence of well-positioned error-correcting repeaters is required for
long-range communication. Quantum networks are particularly affected
by such deterioration \cite{Coutinho2022}. Similarly to the classical
case, the communication range may
be extended using well-placed quantum repeaters \cite{zwerger2018long,Pirandola2019}.
For the purposes of quantum enhanced secure communication, one can
circumvent the need for quantum repeaters using hybrid
classical-quantum networks with trusted nodes \cite{Chen2021}. In such
networks there are some classical repeaters trusted to be secure, and
one only needs to create a quantum connection from one of those
repeaters to another.  In processes of this kind, one may think of
repeaters or trusted nodes as \emph{active}, and long-distance
communication may be possible even if there are no uninterrupted
paths, provided active nodes along the paths are not too far apart.
This calls for a theory of percolation that incorporates a
connectivity definition allowing interruptions/gaps (sequences of
inactive nodes) in the paths between nodes of the same connected
component.  We note that similar phenomena are also found in pure
state quantum networks
\cite{perseguers2008entanglement,Perseguers2010,Xiangyi2021} with
non-trivial entanglement distribution that cannot be tackled using
standard percolation.

Here we propose a general model of extended-range percolation on
complex networks, aiming to provide a mathematical basis for
information transmission involving path interruptions.
Similar problems have been considered in the literature, although, to our knowledge,
exclusively on lattices. In the context of magnetic systems, tunable interaction
ranges were studied in the equivalent neighbour model of Domb, Dalton and Sykes \cite{dalton1964dependence, domb1966crystal}.
Numerous results exist on long-range percolation models where distant nodes of a lattice
are connected with a decaying probability \cite{schulman1983long, aizenman1986discontinuity, biskup2019sharp}. More recent works include
``extended neighbour percolation'' \cite{Malarz2015, malarz2020site, malarz2021percolation, xun2020bond, Xun2021, zhao2022site, xun2022site} and the related ``range-R'' model in mathematics \cite{frei2016lower, hong2023upper}.

Surprisingly, extended-range percolation has not received any attention within the field of
complex networks, thus an understanding of such problems on architectures that may realistically
represent many modern information networks is lacking.

\section{Extended-range percolation model}
\label{sec2}

\begin{figure}[h!]
\centering \includegraphics[width=0.8\columnwidth,angle=0.]{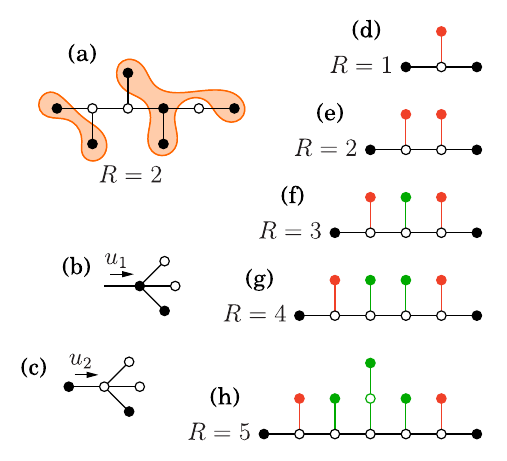}
\caption{Panel (a): A small network with two distinct extended-range connected components, shaded orange.
  Active nodes (filled black circles) have a range of $R=2$. Inactive nodes are represented by empty circles.
  Panels (b) and (c) are schematic representations of Eqs. (\ref{eq10})
  and (\ref{eq20}), respectively.
  Panels (d)-(h) illustrate the concept of off-path bridge nodes
  and the increasing complexity of the problem for larger values of the range
  $R$. Empty nodes are inactive, filled nodes are active.
  The two active black nodes are not directly connected, being at distance
  $R+1$. The red nodes, even if active, do not connect the black nodes indirectly.
  However, for $R>2$, any of the green nodes, which are not on the path
  connecting the two black nodes, play the role
  of a bridge, indirectly connecting them.}
\label{fig:schematic1}
\end{figure}

In an arbitrary
network, let each node be active independently with probability $\phi$,
be inactive otherwise.
Each active node $i$ is able to transmit information
\emph{directly} up to a topological distance $R$, that is, to all
active nodes at most $R$ steps away from node $i$ in the network.
By direct transmission we mean transmission without the help of any other
active nodes potentially acting as repeaters.
We define the \emph{extended-range connected component} of node $i$ as the set of active nodes
that node $i$ is able to transmit information to or receive information from, either directly or
indirectly, via relays of intermediate active nodes.
(Note that $R=1$ corresponds to standard nearest neighbour percolation.)
Fig.~\ref{fig:schematic1}(a) shows an
example of a small network with two extended-range connected components, where active nodes have a range $R=2$.
A perhaps more instructive definition may be given as follows.
Two nodes belong
to the same extended-range connected component if there is at least
one walk between them in which there are at most $R-1$ consecutive
inactive nodes. Note that we must use ''walks" instead of ''paths''
in the definition. This is because two nodes may be indirectly connected
via a third intermediate node which is not on the path between
them. As an example, see Fig. \ref{fig:schematic1}(f) where two active
black nodes are shown on a chain, at a distance of four steps from each other,
therefore not being directly connected for range parameter $R=3$.
An additional off-path active (bridge) node, however, marked in green,
is able to indirectly connect the two nodes,
so that they belong to the same extended-range connected component.
In contrast [Fig.~\ref{fig:schematic1}(d),(e)] off-path bridges
do not allow indirect connections for $R<3$.
This feature introduces a complex
combinatorial problem for high values of $R$, as we detail below.
Here we derive the exact solution of extended-range percolation
on infinite random uncorrelated networks of arbitrary degree distribution.
We start with $R=2$, where off-path bridges do not play
any role, we then tackle the additional complexity due to off-path bridges
for $R=3$ and $4$, then we further generalize to any $R$.
We also present an efficient
message-passing formulation of the theory that works well in finite
real-world networks.

\section{Exact solution in uncorrelated networks}
\label{sec3}

Let us consider a random uncorrelated network with an arbitrary degree distribution $p_k$,
in the infinite size limit.
We start by solving the problem of finding the size of the
extended-range giant connected component (EGCC) for $R=2$.
Let $u_1$ be the probability that, following a random link with
an active end node, we are not able to reach the EGCC
via walks that have gaps of at most one consecutive inactive node.
Probability $u_1$ is suggestively written as
$u_1 = P_{\rightarrow \newmoon}$.
Note that the starting node's state does not play any role.
Let $u_2$ be the probability that following a random link with an
inactive end node and active starting node, we are
not able to reach the EGCC via allowed
walks: $u_2 = P_{\newmoon \rightarrow \fullmoon}$.
A simple equation for $u_1$ is written by noting that
[see Fig.~\ref{fig:schematic1}(b)]
after reaching an active node, the branch is finite only if
all $r$ outgoing neighbors are either
active but do not lead to the EGCC (prob. $\phi u_1$) or
inactive and not leading
to the EGCC (prob. $(1-\phi) u_2$).
Exploiting the local treelikeness of the networks, we can write 

{
\medmuskip=0mu
\thinmuskip=0mu
\thickmuskip=0mu
\begin{align}
u_1 = \sum_{r=0}^{\infty} q_r \big(   \phi u_1 + (1-\phi)u_2    \big)^r = g_1 \big( \phi u_1 + (1-\phi)u_2   \big),  \label{eq10}
\end{align}
}

\noindent
where $q_r$ is the excess degree distribution and $g_1$ is its
probability generating function.
The equation for the probability $u_2$ is similar, but in this case,
having arrived from an active node to an
inactive one, any inactive outgoing neighbor surely does not lead to
the EGCC (see Fig.~\ref{fig:schematic1}(c)),
{
\medmuskip=0mu
\thinmuskip=0mu
\thickmuskip=0mu
\begin{align}
u_2 = g_1 \big( \phi u_1 + (1-\phi)   \big).  \label{eq20}
\end{align}
}

\noindent
The relative size of the EGCC (probability of a randomly chosen node
belonging to the EGCC) is
\medmuskip=0mu
\thinmuskip=0mu
\thickmuskip=0mu
\begin{align}
S = \phi \Big[ 1 - g_0 \Big( \phi u_1 + (1-\phi) u_2  \Big)  \Big],  \label{eq55}
\end{align}

\noindent
where $g_0$ is the generating function for the degree distribution $p_k$.
Setting $u_2=1$ leads back to standard nearest-neighbor percolation.

Extending the theory to $R=3$ and $R=4$ requires the introduction of
two other probabilities. Let us define $u_3$
as the probability that, following a random link with an inactive end
node, and an inactive starting node, but with at least one active
neighbor of the starting node, we are not able to reach the EGCC
via allowed walks:
$u_3 = P_{\newmoon \relbar \fullmoon \rightarrow \fullmoon}$.
Finally, $u_4$ is
the probability that we are not able to reach the EGCC
via allowed walks, by following a link that has an
inactive end node, and has at least one active node at a distance 3 in
the ``reverse direction'', but no active nodes at shorter distances:
$u_4 = P_{\newmoon \relbar \fullmoon \relbar \fullmoon \rightarrow \fullmoon}$.
In the case $2<R\le 4$ the equation for $u_1$ remains the same, while
the equation for $u_2$ is trivially modified, as one has to take into
account that the EGCC could be reached, with probability $u_3$,
even if the neighbor of an
inactive node is inactive [see Fig.~\ref{fig:schematic1}(c)],
\medmuskip=0mu
\thinmuskip=0mu
\thickmuskip=0mu
\begin{align}
u_2 = g_1 \big( \phi u_1 + (1-\phi) u_3  \big).  
\label{eq20bis}
\end{align}
The equation for $u_3$ is more complicated
because it may happen 
[see Fig.~\ref{fig:schematic1}(f) for the case of $R=3$] that two nodes belong to the same component
even if there are three inactive nodes along the path between them.
The indirect connection is guaranteed by the presence of an active off-path bridge node
[the green node in Fig.~\ref{fig:schematic1}(f)] which is at a distance $3$
from both of them. [The same phenomenon is shown in Fig. \ref{fig:schematic1}(g) for $R=4$.]
Taking explicitly into account this possibility, for $R=4$ the
equations for $u_3$ and $u_4$ are (see Appendix \ref{app1} for details)
\medmuskip=0mu
\thinmuskip=0mu
\thickmuskip=-0.2mu
\begin{align}
\label{eq40}
u_3&=g_1\big(  (1-\phi) u_4 \big) + g_1 \big( \phi u_1 + (1-\phi) u_3 \big) - g_1 \big(  (1-\phi) u_3 \big),\\
\label{eq50}
u_4&= g_1(1-\phi) + g_1 \big( \phi u_1 + (1-\phi) u_3 \big) - g_1 \big( (1-\phi) u_3 \big).
\end{align}
For $R=3$, Eq.~\eqref{eq50} is simply replaced by $u_4=1$.

The extension of the approach to $R>4$ is highly nontrivial, as it
requires the introduction of increasingly complex conditional probabilities.
Physically, this is due to the fact that bridges involve
off-path nodes at increasing distances from the path
[Fig.~\ref{fig:schematic1}(h)].
See the Appendix for a presentation of the framework valid for any $R$
and the explicit equations for $R=5$ and $R=6$.

\begin{figure}[H]
\centering \includegraphics[width=\columnwidth,angle=0.]{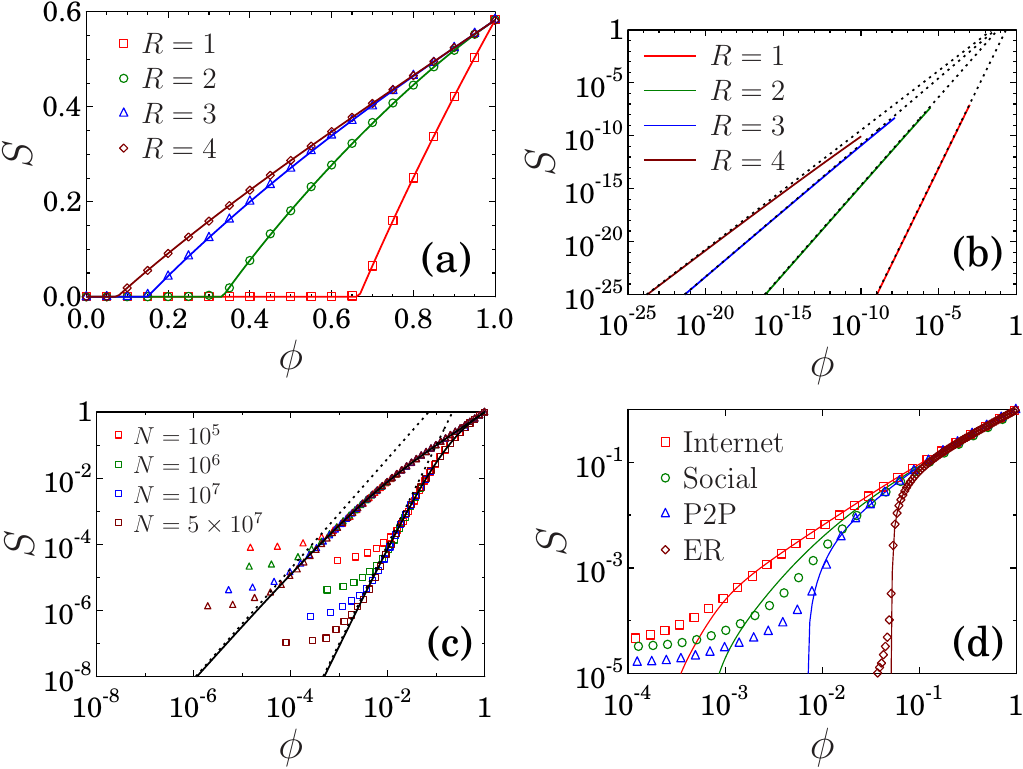}
\caption{Relative size $S$ of the EGCC as a function of node activation
  probability $\phi$. (a) Predictions obtained solving numerically the
  equations for the $u_i$ (solid lines) compared with simulation results
  (symbols) for an Erd\H os-R\'enyi network of $\langle k \rangle = 1.5$, for
  $R=1,\ldots,4$.
  (b) Numerical solutions of the same equations (solid lines) compared
  with the predicted singular behavior (dashed black lines) for power-law networks
  [Eqs.~\eqref{eq:new_critical_exponent}] with $\gamma=2.5$ for $R=1,\ldots,4$.
  (c) Comparison of the numerical solutions of Eqs. (\ref{eq10}),
  (\ref{eq20}) and (\ref{eq55}) (solid black lines)
  with simulation results for $\gamma=2.5$, $R=1$ (squares)
  and $R=2$ (triangles), for different network sizes.
  The dashed black lines correspond to the predicted singular behaviour [see
  Eqs.~\eqref{eq:new_critical_exponent} for $R=2$ and Ref. \cite{radicchi2015breaking} for the standard $R=1$ case].
  Analogous results for $\gamma>3$ are shown in Appendix \ref{app4}.
  (d) Solutions of the message-passing
  equations (\ref{eq90}), (\ref{eq100}) and (\ref{eq110}) (solid lines) compared with
  simulation results (symbols) for three example real-world networks and an
  Erd\H os-R\'enyi network of $\langle k \rangle = 4$.}
\label{fig:together}
\end{figure}

Our approach is exact for random uncorrelated locally tree-like
networks with an arbitrary degree distribution.  To find the solution,
up to arbitrary precision, the equations for the probabilities $u_i$
may be iterated until convergence.  Together with Eq.~\eqref{eq55}
they allow to determine the size of the EGCC.
Fig.~\ref{fig:together}(a) displays the behavior of $S$ calculated
using the numerical solutions of the equations for the probabilities $u_i$,
for an Erd\H os-R\'enyi network of mean degree
$\langle k \rangle = 1.5$, for $R=1,\ldots,4$.
Simulation results are also shown
and demonstrate perfect agreement with the theoretical predictions.
Equally perfect agreement
is shown in the Appendix for power-law distributed networks.
As expected, the size of the EGCC is zero up to a percolation threshold
$\phi_c^{(R)}$, which gets smaller as $R$ is increased.

To determine the critical threshold, let us consider
for simplicity the case $R=4$.
Eqs. (\ref{eq10}), (\ref{eq20bis}), (\ref{eq40}) and
(\ref{eq50}) have the trivial solution $u_1=u_2=u_3=u_4=1$ for any value
of $\phi$, which corresponds to $S=0$. This solution becomes
unstable at the critical threshold $\phi_c^{(R=4)}$, above which the only
stable solution corresponds to $S>0$. To find this threshold, we study the
stability of the trivial solution by linearizing the equations around it, to
obtain the Jacobian matrix

\begin{align}
\hat{J} = b
\begin{pmatrix}
 \phi & 1-\phi & 0 & 0 \\
 \phi & 0 & 1-\phi & 0 \\
 \phi & 0 & \frac{(1-\phi)[ b - g_1'(1-\phi) ]}{b} & \frac{(1-\phi) g_1'(1-\phi)}{b} \\
 \phi & 0 & \frac{(1-\phi)[ b - g_1'(1-\phi) ]}{b} & 0
\end{pmatrix},
\label{eq70}
\end{align}

\noindent
where $b$ is the mean branching, $b=\langle k(k-1) \rangle / \langle k \rangle$.
Using the Perron-Frobenius theorem, the critical threshold occurs when $\lambda_1$,
the largest real eigenvalue of $\hat{J}$ is equal to $1$.
Thus the critical condition is $\det[\hat{J} - \hat{I}] = 0$, where
$\hat{I}$ is the identity matrix.
For $R<4$ the same methodology applies, only we must consider the
appropriate smaller matrix of dimensions $R \times R$ in the top left
corner of matrix $\hat{J}$. In this way we recover the standard
percolation threshold ($R=1$) and a simple formula for the threshold
in the case of $R=2$,
\begin{align}
\phi_c^{(R=1)} = \frac{1}{b}, \quad \quad \phi_c^{(R=2)} = \frac{1+b-\sqrt{ (1+b)^2 - 4 }}{2b}.  \label{eq80}
\end{align}

\noindent
Analyzing the asymptotics of $\phi_c^{(R=2)}$ we obtain that
$\phi_c^{(R=2)} \to b^{-2}$ for $b \to \infty$.
For $R=3$ and $R=4$ the critical condition does not reduce
to an algebraic equation for $\phi_c^{(R)}$, but may be solved easily by numerical means.

By expanding the r.h.s. of Eqs. (\ref{eq10}), (\ref{eq20bis}),
(\ref{eq40}), (\ref{eq50}) and (\ref{eq55}) in powers of $\epsilon_i=1-u_i$ one
can show (see Appendix \ref{app4}) that, close to the threshold, the size of the EGCC behaves
as $S \sim (\phi-\phi_c^{(R)})^\beta$, with $\beta = 1$ for
non-heterogeneous networks, i.e. networks whose degree distributions
have a finite third moment, which includes ER networks and power-law
degree-distributed networks, $p_k\sim k^{-\gamma}$, with $\gamma>4$.
For weakly heterogeneous networks,
with $3<\gamma<4$, we find the nontrivial exponent $\beta=1/(\gamma-3)$
as in ordinary percolation \cite{radicchi2015breaking}. 
For strongly heterogeneous networks, with $2<\gamma<3$,
we find a non-universal behavior, with
a critical exponent depending both on $\gamma$ and $R$.
In particular, $S \sim \phi^{\beta}$, with
\medmuskip=0mu
\thinmuskip=0mu
\thickmuskip=0mu
\begin{eqnarray}
  \label{eq:new_critical_exponent}
  \nonumber
\beta = & 1 +\frac{\gamma-2}{1-(\gamma-2)^2}&~~~~ \textrm{for }R=2,\\
\beta = & 1 +(\gamma-2)^{R-1} &~~~~ \textrm{for }R > 2.
\end{eqnarray}
Panels (b) and (c) of Fig.~\eqref{fig:together} display the $R$-dependence
of the critical behavior of $S$ and show, for $R=2$, that numerical
simulations agree with the predicted behavior.
The dependence on $R$ may be surprising at first glance, since the
present extended-range percolation model is not ``long-range''
(interaction ranges are finite) as opposed to truly
long-range percolation~\cite{Cirigliano2022}.
The nonuniversality stems from the fact that for strongly
heterogeneous networks ($\gamma<3$), the average number of second
(and third, fourth, etc.) neighbors is infinite,
hence each node is able to interact with an infinite number of peers.

\section{Message-passing equations}
\label{sec4}

Based on Eqs. (\ref{eq10}), (\ref{eq20bis}), (\ref{eq40}), (\ref{eq50})
and (\ref{eq55}), it is straightforward to construct a set of
message-passing equations for the case of $R\leq4$, to allow for an
approximate solution of extended-range percolation on non-random networks,
specified by a given adjacency matrix.
Here we present the special case of $R \leq 2$, the
general description may be found in Appendix \ref{app5}.
For $R=2$ the message-passing equations corresponding to the
self-consistency equations (\ref{eq10}) and (\ref{eq20}) are written
as
\medmuskip=0mu
\thinmuskip=0mu
\thickmuskip=0mu
\begin{align}
u^{(i \leftarrow j)}_1 &= \prod_{k \in \partial_j \setminus i} \left[ \phi u^{(j \leftarrow k)}_1 + (1-\phi) u^{(j \leftarrow k)}_2  \right],  \label{eq90}  \\
u^{(i \leftarrow j)}_2 &= \prod_{k \in \partial_j \setminus i} \left[ \phi u^{(j \leftarrow k)}_1 + 1-\phi  \right],  \label{eq100}
\end{align}
where $\partial_j$ denotes the set of neighbors of node $j$.
The quantity $u^{(i \leftarrow j)}_1$ is the probability that
following link $(ij)$, in the direction of $j$, we are not able to
reach the EGCC via walks that have gaps of
at most one inactive node, given that node $j$ is active.
Similarly, $u^{(i \leftarrow j)}_2$ is the probability that
following link $(ij)$, in the direction of $j$, we are not able to
reach the EGCC via walks that have gaps of
at most one inactive node, given that node $j$ is inactive
and node $i$ is active.
Eqs. (\ref{eq90}) and (\ref{eq100}) are asymptotically exact (in the infinite size limit) in locally
treelike networks.
The probability that node $i$ belongs to the EGCC can then be expressed as
\medmuskip=0mu 
\thinmuskip=0mu
\thickmuskip=0mu
\begin{align}
S_i = \phi \left[ 1- \prod_{j \in \partial_i} \left( \phi u^{(i \leftarrow j)}_1 + (1-\phi) u^{(i \leftarrow j)}_2  \right) \right],  \label{eq110}
\end{align}

\noindent
and the relative size of the EGCC as the average $S = N^{-1} \sum_{i=1}^N S_i$,
where $N$ is the number of nodes. Fig. \ref{fig:together}(d) shows
message-passing results compared with simulation results, for an Erd\H
os-R\'enyi network and three example real-world networks (see Appendix \ref{app6}
for details).
The correspondence is perfect in the case of Erd\H
os-R\'enyi, as expected, and also very good for the real-world
networks containing short loops. In the case of the network ``Social'' the fit is somewhat poorer, due to the very high average clustering coefficient in this network, $C\approx 0.65$.

Within the message-passing approach the critical threshold $\phi_c^{(R)}$
may be obtained by following the standard
procedure, i.e., linearizing Eqs. (\ref{eq90}), (\ref{eq100})
around the trivial solution, $u^{(i \leftarrow j)}_1 = u^{(i \leftarrow j)}_2 = 1$
for all links $(ij)$. We obtain, for $R=2$, the Jacobian matrix
\medmuskip=0mu
\thinmuskip=0mu
\thickmuskip=0mu
\begin{align}
\hat{C} =
\begin{pmatrix}
 \phi \hat{B} & (1-\phi) \hat{B} \\
 \phi \hat{B} & 0
\end{pmatrix},
\label{eq120}
\end{align}

\noindent
where $\hat{B}$ is the nonbacktracking (or Hashimoto) matrix
\cite{krzakala2013spectral}.
The largest eigenvalue of matrix $\hat{C}$ must be $1$ at the
threshold.
The solution may be found numerically;
e.g., Ref.~\cite{baxter2021degree} provides an adaptation of the
Newton-Raphson method for such problems, whereby one can quickly find $\phi_c^{(R=2)}$
to arbitrary precision.
When $R=1$, we
have $\hat{C} = \phi \hat{B}$, and the problem reduces to the
message-passing formulation of ordinary percolation \cite{Karrer2014}.

\section{Non-uniform range parameter}
\label{sec5}

\begin{figure}[h!]
\centering \includegraphics[width=0.75\columnwidth,angle=0.]{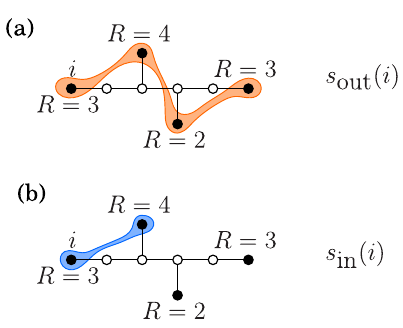}
\caption{Extended-range out-component (a) and in-component (b) of node $i$. The range of each active node (filled black circles) is indicated near it.}
\label{fig:schematic2}
\end{figure}

One may consider a more general version of the model presented above, where the range parameter
of each node is allowed to be different.
Specifically, let each active node $i$ be able to transmit information
directly up to a topological distance $R_i$.
We define the \emph{extended-range
  out-component} $s_{\textrm{out}}(i)$ of node $i$ as the set of active nodes to which node
$i$ is able to transmit information, either directly or indirectly,
via relays of intermediate active nodes.
Fig.~\ref{fig:schematic2}(a) shows an
example of a small network with the extended-range out-component of
node $i$ shaded orange. Note that node $i$ is only able to directly
transmit information to one active node, while two further active nodes are
reached indirectly. The \emph{extended-range in-component} $s_{\textrm{in}}(i)$ of node
$i$ is the set of all active nodes that are able to transmit information to
node $i$, either directly or indirectly [see Fig. \ref{fig:schematic2}(b), shaded blue].
In the case of uniform range parameter ($R_i = R$) the extended-range out- and in-components of
any node coincide and are equivalent to the extended-range connected component, defined in Sec. \ref{sec2}.

It is interesting to note that the non-uniform extended-range percolation model on any network is equivalent to
standard directed percolation on a modified network where we add directed links pointing from all nodes $i$ to
all other nodes at most a distance $R_i$ from node $i$. (The original links should be considered undirected, or
``bidirectional''.) As a result, the concept of strong connectivity, as well as a rich variety of topologically
different components emerge, as in standard directed percolation \cite{broder2000graph,timar2017mapping}.

\section{Discussion and conclusions}
\label{sec6}

We have presented a framework for solving
extended-range percolation exactly on locally tree-like networks.
This generalization of ordinary percolation, allowing for gaps in the walks
connecting distant nodes, provides a fundamental topological
description of connectivity properties in noisy quantum networks,
hybrid classical-quantum networks with trusted nodes, or classical
data transmission networks using error-correcting repeaters.
Our results, combined with existing methods of optimal
percolation~\cite{morone2015influence, morone2016collective, braunstein2016network}
may suggest possible strategies for the optimal placement of repeaters (or trusted
nodes) in such communication networks.

Other interesting perspectives are opened by our work.
The approach
presented here may be used and extended to further analyze the
critical properties of the extended-range percolation transition in
networks, beyond the derivation of the exponent $\beta$, including in
particular finite-size scaling and the distribution of finite
extended-range connected components~\cite{Cirigliano2023a}.

Extended-range percolation on a generic undirected graph
$\mathcal{G}_0^{(u)}$ can actually be seen, if all $R_i=R$,
as a standard nearest-neighbor percolation process on a different
graph $\mathcal{G}_1^{(u)}$, obtained from the
original one by adding
a link from each node $i$ to all other nodes at most $R$ steps away from it.
\footnote{
A similar percolation problem on an ``infinite dimensional'' substrate was considered
in Ref. \cite{vspakulova2009critical}, where the bond percolation threshold of a
specially constructed ``grandparent tree'' was derived. This tree is obtained by adding links
between nodes and their grandparent (relative to a root) in a Cayley tree, thus introducing loops
of length three and destroying the tree structure. Importantly, however, in this construction links
between different children of a node are not added, making it a distinct problem from the one considered here.
}
The graph $\mathcal{G}_1^{(u)}$ is clearly highly clustered.
Hence our approach, which exactly solves extended-range percolation
on the original tree-like graph, provides also the exact solution
for ordinary percolation on the corresponding graph with many intertwined short loops.
This is a rare case of a highly clustered network model
whose nontrivial percolation properties are exactly known.
The formation of the $\mathcal{G}_1^{(u)}$ clustered graph by the
addition of links among neighbors at short distances, can be seen as a
variant of the triadic closure mechanism, prevalent in many real-world
networks \cite{asikainen2020cumulative}. Our results may therefore
also aid the development of more precise theories of percolation and
related processes (e.g., epidemic spreading) in such cases.
Exploring the topological properties of $\mathcal{G}_1^{(u)}$ \cite{Cirigliano2023b} and
of the directed graphs $\mathcal{G}_1^{(d)}$ arising in the case
of non-uniform range parameters $R_i$, presents an additional
interesting avenue for future research.

Another intriguing aspect of extended-range percolation is the question of how site and
bond percolation are related in this problem. We have considered site percolation in
this work, for which it is true that extended-range percolation on a graph $\mathcal{G}_0^{(u)}$
is equivalent to standard percolation on the modified graph $\mathcal{G}_1^{(u)}$, defined above.
The same does not hold for bond percolation, however. It would be interesting to compare the
critical thresholds and critical exponents of the two types of processes, and to find an
interpretation of extended-range bond percolation in terms of the modified graph $\mathcal{G}_1^{(u)}$.

\section*{Acknowledgments}

We are grateful to Bruno Coutinho for useful discussions.  This work
was developed within the scope of the project i3N, UIDB/50025/2020 \&
UIDP/50025/2020, financed by national funds through the
FCT/MEC--Portuguese Foundation for Science and Technology. G.T. was
supported by FCT Grant No. CEECIND/03838/2017.

\begin{widetext}


\appendix

\section{Derivation of the equations in the case $R=4$}
\label{app1}

In this section we present the derivation of  Eqs.~$(5)$ and $(6)$ in the main text.
We refer to the main text for the definition of the probabilities
$u_1 = P_{\rightarrow \newmoon}$,
$u_2 = P_{\newmoon \rightarrow \fullmoon}$,
$u_3 = P_{\newmoon \relbar \fullmoon \rightarrow \fullmoon}$
and $u_4 = P_{\newmoon \relbar \fullmoon \relbar \fullmoon \rightarrow
\fullmoon}$ and for
the derivation of the equations for $u_1$ and $u_2$.
Concerning $u_3$, let us look at Fig.\ref{fig:drawing}
%
%
and consider node $i$, the node at which we arrive following the configuration
that corresponds to $u_3$. Let the number of outgoing neighbors, i.e., the excess
degree at node $i$, be denoted by $r$.
If none of the outgoing neighbors of node $i$
are active [Fig.~\ref{fig:drawing}(a)], an event
occurring with probability $(1-\phi)^r$,
the EGCC is not reached provided none of them leads to it, an event
that occurs with probability $u_4$ for each of the independent branches.
If instead there are $n>0$ active nodes among the $r$ outgoing neighbors
[Fig.~\ref{fig:drawing}(b)], the probability of not reaching the EGCC is
$(\phi u_1)^{n}[(1-\phi)u_3]^{r-n}$.
Note that in this case the branches are no longer independent, as the
probability (of not reaching the EGCC) on a branch leading to an inactive node is modified by the
presence of an active node in another branch, potentially acting as an off-path bridge node.
Since only the number of active neighbors matters, and not where they are
placed, ${r \choose n}$ configurations give the same contribution to $u_3$.
Summing over $n$ we obtain
\begin{align}
\nonumber
& [(1-\phi)u_4]^r +\sum_{n=1}^{r} {r \choose n}(\phi u_1)^{n}[(1-\phi)u_3]^{r-n}\\ \nonumber
&= [(1-\phi)u_4]^r -[(1-\phi)u_3]^r+ \sum_{n=0}^{r} {r \choose n}(\phi u_1)^{n}[(1-\phi)u_3]^{r-n} \\ \nonumber
&= [(1-\phi)u_4]^r -[(1-\phi)u_3]^r+ \big(\phi u_1+(1-\phi)u_3\big)^r.
\end{align}
Finally, averaging over the excess degree distribution $q_r$ we get
\begin{align}
\nonumber
u_3 &= \sum_r q_r\left\{[(1-\phi)u_4]^r -[(1-\phi)u_3]^r+
    \big(\phi u_1+(1-\phi)u_3\big)^r  \right\} \\
&= g_1\big((1-\phi)u_4\big)-g_1\big((1-\phi)u_3\big)+g_1(\phi u_1 +(1-\phi)u_3),
\label{eq:u_3}
\end{align}
which is Eq.~(5) in the main text.

To derive $u_4$, the argument is perfectly analogous.
If we arrive at node $i$ from a configuration
${\newmoon \relbar \fullmoon \relbar \fullmoon \rightarrow \fullmoon}$,
then if none of the outgoing neighbors of node $i$ are active
[Fig.~\ref{fig:drawing}(c)] the EGCC is not reached.
This occurs with probability $(1-\phi)^r$.
On the contrary, if at least one of the outgoing neighbors is active
[Fig.~\ref{fig:drawing}(d)] then the probability of not reaching the EGCC on branches leading to inactive nodes is modified.
Following the same procedure as above we get
\begin{equation}
u_4 = g_1(1-\phi)-g_1\big((1-\phi)u_3\big)+g_1(\phi u_1 +(1-\phi)u_3),
\label{eq:u_4}
\end{equation}
which is Eq.~$(6)$ in the main text.

These equations are valid for $R=4$.
The equations for $R=3$ are obtained by simply setting $u_4=1$.
\begin{figure}
\includegraphics[scale=0.8]{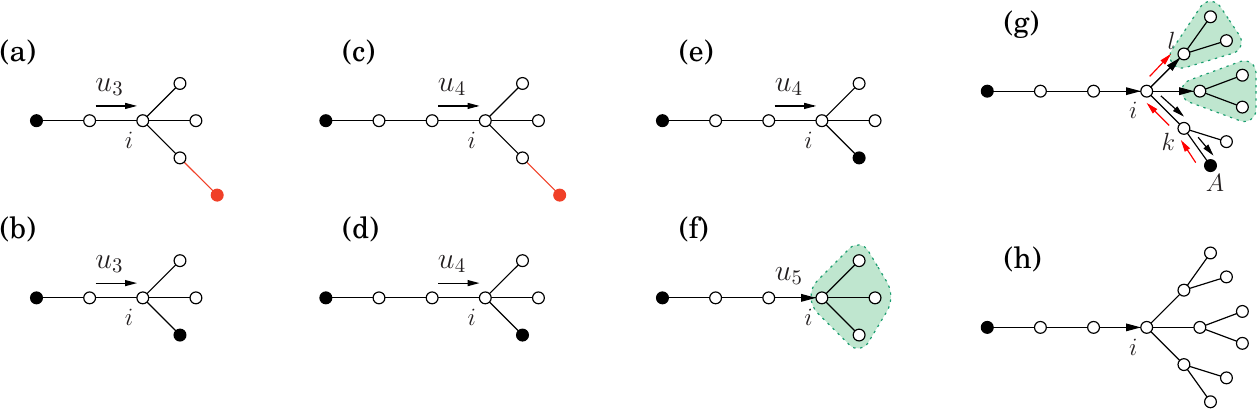}
\caption{Visual representation of the recursive equations for $u_3$,
  $u_4$ and $u_5$.
  Panels (a)-(d) are for the case $R=4$, while panels
  (e)-(h) are for $R=5$. In (a) and (b) we have the two possible
  scenarios contributing to $u_3$. It is important to note that even
  if a second neighbor of node $i$ is active [the red node in panel
  (a)], the probability along the other branches still remains
  $(1-\phi)u_4$. In (c) and (d) the configurations contributing to
  $u_4$ are shown. Panels (e) and (f) show how the equations for $u_4$
  must be modified for $R=5$ to take into account correlations among
  branches. Correlations due to off-path bridge nodes at distance $2$
  are contained in the probability $u_5$.  Panels (g) and (h) show the
  two possible scenarios contributing to $u_5$. In particular, from
  panels (g) and (h) we see that the probability of not reaching the
  EGCC via the configuration associated with node $l$ is different
  depending on whether node A, which belongs to another branch, is
  active or inactive.
  }
\label{fig:drawing}
\end{figure}
\section{The equations for $R=5$}
\label{app2}

In the case $R=5$ the combinatorial difficulty of the problem
is strongly increased, because
one has to take into account dependencies between different branches
arising due to off-path bridge nodes that are at distance 2 from
the focal node $i$.
A first observation is that the equation for $u_3$
remains the same as in the $R=4$ case, Eq.~\eqref{eq:u_3}.
This happens because if none of the outgoing neighbors of node $i$
are active, then even if we can reach an active
node at distance $2$ from $i$, this does not change the probability
along the other branches, which remains $u_4$,
as one node at distance $4$
[the leftmost in Fig.~\ref{fig:drawing}(a)] is surely active.
Hence in this case we can
conclude that the branches are independent and the corresponding
probability is $[(1-\phi)u_4]^r$.
A similar argument implies that also the equations for $u_1$ and $u_2$ are unchanged.

Let us consider now the configuration
${\newmoon \relbar \fullmoon \relbar \fullmoon \rightarrow \fullmoon}$
corresponding to $u_4$ and let us call again $i$ the node at which we arrive.
We must now consider the state of neighbors up to distance $2$, and the presence of an active
second-neighbor node along one branch changes the probability
of not reaching the EGCC along the others.
We can start by distinguishing two complementary scenarios:
\begin{enumerate}
\item $n>0$ nodes of $i$'s first neighbors are active [Fig.~\ref{fig:drawing}(e)],
  i.e., there is at least one off-path bridge node at distance 1.
  In such a case we can repeat the argument presented above to derive Eq.~\eqref{eq:u_3} and
  conclude that the probability of not reaching the EGCC is
  \begin{equation}
    v^{(r)}=\big(\phi u_1 +(1-\phi)u_3\big)^r-((1-\phi)u_3)^r;
    \label{simple}
  \end{equation}
\item none of $i$'s first neighbors are active [Fig.\ref{fig:drawing}(f)].
  In this case we denote the probability of not reaching the EGCC by $u^{(r)}_5$.
\end{enumerate}
Hence we can write, averaging over the excess degree $r$,
\begin{equation}
u_4 = \sum_{r} q_r \left(u^{(r)}_5+v^{(r)} \right) =  u_5 + v = u_5 -g_1\big((1-\phi)u_3\big)+g_1(\phi u_1 +(1-\phi)u_3).
\end{equation}
We still need an equation for the probability $u_5$.
We stress that the meaning of $u_5$ is conceptually different from the
previous $u_i$ defined for $i \le 4$.
Due to the possibility of having off-path bridge nodes at distance 2,
the branches emanating from the focal point $i$ are not independent and
$u_5$ is a probability associated with the state of all
neighbors of node $i$.
In particular, this means that $u_5 \neq P_{\newmoon \relbar \fullmoon \relbar \fullmoon \relbar \fullmoon \rightarrow \fullmoon}$.
Instead $u_5$ is defined as the
probability of not reaching the EGCC following a link which
ends up in a configuration consisting of an inactive node with all the
outgoing neighbors inactive, arriving from a branch of the type
${\newmoon \relbar \fullmoon \relbar \fullmoon \rightarrow \fullmoon}$,
i.e. knowing that an active node is at distance $3$ in
the reverse direction [see Fig.~\ref{fig:drawing}(f), where this
configuration of only inactive outgoing neighbors is shaded green].
Thus $u_5$ is the probability that the configuration in Fig.~\ref{fig:drawing}(f) does not lead to the EGCC.
If we arrive at the configuration associated with $u^{(r)}_5$---that is, with $r$ empty outgoing neighbors---, then we must consider the state of nodes at distance $2$ from $i$ to calculate the probability that the EGCC is not reached.

Again, two complementary scenarios must be considered:
\begin{enumerate}
\item $n>0$ of the branches lead to an active second neighbor of node $i$ [Fig.\ref{fig:drawing}(g)], i.e., there is at least one
  off-path bridge node at distance 2.
\item none of the branches lead to an active second neighbor [Fig.\ref{fig:drawing}(h)].
  \end{enumerate}

Let us analyze the case (a).
Consider a first neighbor $k$ of node $i$, with excess degree $\rho_k$ and $\tau_k \ge 1$ of its outgoing neighbors active.
The probability of not reaching the EGCC through it is
\begin{equation}
\sum_{\tau_k=1}^{\rho_k}{\rho_k \choose \tau_k}(\phi u_1)^{\tau_k}((1-\phi)u_3)^{\rho_k-\tau_k}=v^{(\rho_k)}.
\end{equation}
For a first neighbor $l$ of node $i$ with none of the other neighbors active,
the probability of not reaching the EGCC through it is instead $u^{(\rho_l)}_5$. Note that we can use $u^{(\rho_l)}_5$ because node $l$ in Fig.\ref{fig:drawing}(g)
is at distance $3$ from node $A$, which is active, and no other active nodes at smaller distances are present.
Summing over all the possible ways of placing these branches we obtain
\begin{align}
\nonumber
&\sum_{n=1}^{r} {r \choose n} \prod_{k=1}^{n} v^{(\rho_k)}\prod_{l=1}^{r-n}u^{(\rho_l)}_5.
\end{align}
Averaging over the excess degrees of nodes $k$ and $l$ we get
\begin{align}
\sum_{n=1}^{r} {r \choose n} \prod_{k=1}^{n} v \prod_{l=1}^{r-n}u_5 =\sum_{n=1}^{r} {r \choose n} v^n u_5 ^{r-n} =-u_5^r+\left(u_5+v \right)^r.
\label{second}
\end{align}

In the case (b) all second-neighbors of node $i$ are inactive
[Fig.~\ref{fig:drawing}(h)].  This happens with probability
$(1-\phi)^{ \sum_{j=1}^r \rho_j}$, so that averaging over the excess
degrees ${\rho_j}$ we get a contribution $\big(g_1(1-\phi)\big)^r$.
Summing this last contribution with the other in Eq. \eqref{second},
and averaging over the excess degree $r$, considering the overall
multiplicative factor of $(1-\phi)^r$, we obtain
\begin{align}
\nonumber
u_5 &= \sum_{r} q_r(1-\phi)^r\left[\big(g_1(1-\phi)\big)^r -u_5^r+\left(u_5+v\right)^r\right]\\
&= g_1 \big( (1-\phi)g_1(1-\phi)\big)-g_1\big((1-\phi)u_5\big)+g_1\big((1-\phi)(u_5+v) \big).
\label{u_4}
\end{align}
Hence we end up with the two equations for $u_4$ and $u_5$
\begin{align}
\label{eq:u4}
u_4 &= u_5 - g_1\big((1-\phi)u_3\big)+g_1(\phi u_1 +(1-\phi)u_3),\\
u_5 &= g_1 \big( (1-\phi)g_1(1-\phi)\big)-g_1\big((1-\phi)u_5\big)+g_1\big((1-\phi)(u_5- g_1\big((1-\phi)u_3\big)+g_1(\phi u_1 +(1-\phi)u_3)) \big).
\label{eq:w4}
\end{align}
Eqs.~\eqref{eq:u4} and \eqref{eq:w4}, together with the equations for $u_1$, $u_2$ and $u_3$
constitute a set of five recursive equations involving
the five probabilities, $u_1,\dots, u_5$.
Solving them iteratively one can evaluate $S$ using Eq.~(3) in the main text.





\section{The case $R>5$}
\label{app3}

Following the same lines of argument as above, it is
straightforward to derive the equations for $R=6$ (the equations for
$u_i$ up to $i=4$ are unchanged)
\begin{align}
\label{eq:w_4_}
u_5 &= g_1 \big( (1-\phi)u_6 \big)-g_1\big((1-\phi)u_5\big)+g_1\big((1-\phi)(u_5- g_1\big((1-\phi)u_3\big)+g_1(\phi u_1 +(1-\phi)u_3)) \big),\\
\label{eq:w_5}
u_6 &= g_1 \big( (1-\phi)g_1(1-\phi)\big)-g_1\big((1-\phi)u_5\big)+g_1\big((1-\phi)(u_5- g_1\big((1-\phi)u_3\big)+g_1(\phi u_1 +(1-\phi)u_3)) \big).
\end{align}
Here the probability $u_6$ corresponds to a configuration analogous to $u_5$, only in this case we arrive from a configuration where the active node in the reverse direction is one step further away. In other words, $u_6$ is the probability of not reaching the EGCC having arrived from a configuration of type ${\newmoon \relbar \fullmoon \relbar \fullmoon \relbar \fullmoon \rightarrow \fullmoon}$, given that all outgoing neighbors (of the node at which we arrive) are inactive.
Setting $u_6 = g_1(1-\phi)$ we recover the equations for $R=5$.

A similar scheme may be used also for $R>6$.
For any value of $R$ one can set up $R$ equations for $R$ distinct probabilities $u_1,\ldots,u_R$. For increasing $R$, off-path bridge nodes may be situated at larger and larger distances, therefore one must introduce probabilities corresponding to increasingly complex configurations. The probabilities $u_i$ for $i>6$ are analogous to $u_5$ and $u_6$, but are conditioned on all outgoing first, second, third, etc. neighbors being inactive. This would result in increasingly complex equations involving multiply nested
generating functions. Note that the complexity of the configurations associated with the probabilities $u_i$, and consequently, the complexity of the corresponding equations, increases in steps of two, due to the general observation that off-path bridge nodes at distance $r$ can only play a role for $R \geq 2r+1$.

The theoretical predictions for $R=5$ and $R=6$ are tested in
Fig.~\ref{fig:simul}(a) and Fig.~\ref{fig:simul}(b), which display $S(\phi)$
given by Eq.~(3) in the main text and using the numerical solution of
the equations for the probabilities $u_i$, for
$R=1,\dots,6$, for power-law degree-distributed networks ($p_k \sim
  k^{-\gamma}$) with (a) $\gamma=3.5$ and (b) $\gamma=4.5$, respectively.
In both cases, results of numerical simulations are in agreement
with the theoretical predictions.

\begin{figure}[H]
\includegraphics[scale=0.7]{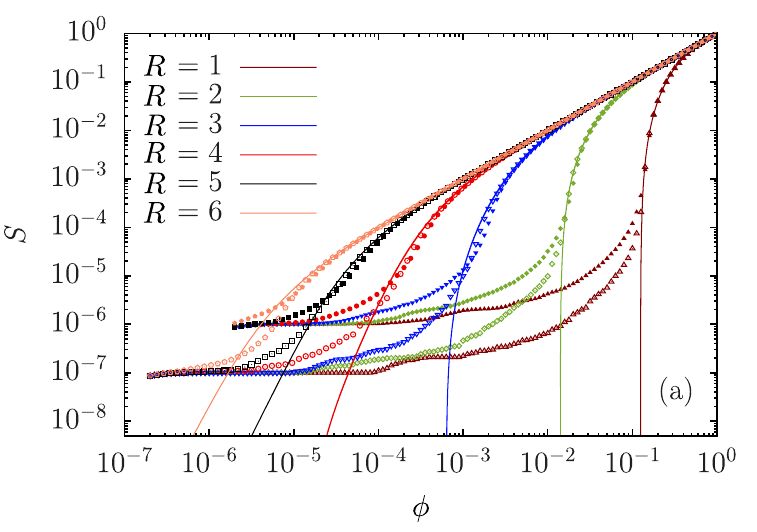}
\includegraphics[scale=0.7]{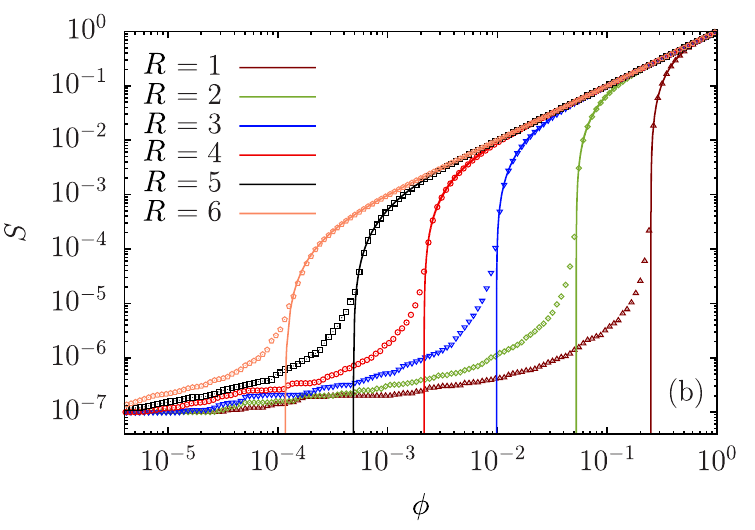}
\includegraphics[scale=0.7]{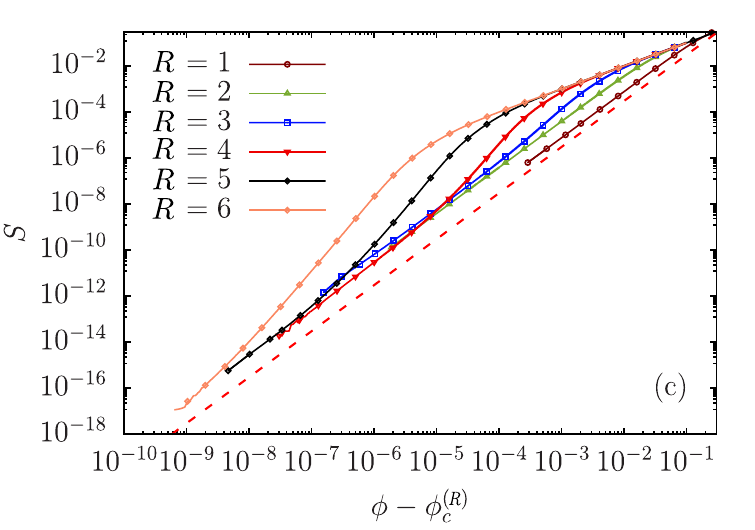}
\includegraphics[scale=0.7]{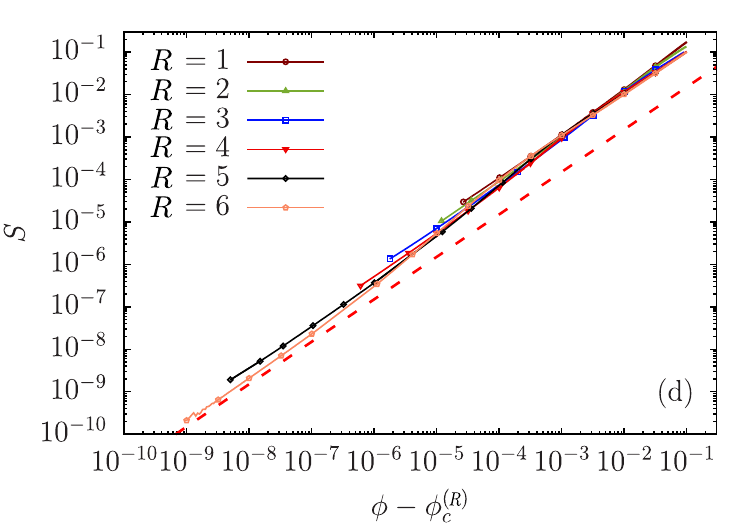}
\caption{Top: comparison between numerical simulations (symbols),
  performed using a modification of the Newman-Ziff algorithm
  \cite{newman2001fast}, and exact solution (lines) obtained solving
  Eq.~(3) in the paper and the recursive equations for the
  probabilities $u_i$ for $R=1,\dots,6$, for (a) $\gamma=3.5$,
  $N=10^6$ (full symbols), $N=10^7$ (empty symbols) and (b)
  $\gamma=4.5$, $N=10^7$, respectively. Bottom: scaling of the exact
  solution for $S$, obtained as before, versus $\delta \phi$, for (c)
  $\gamma=3.5$ and (d) $\gamma=4.5$, respectively. Dashed lines
  represent the scaling for $\delta \phi$ small with the exponent
  $\beta=1/(\gamma-3)$ and $\beta=1$ for figure (c) and (d)
  respectively, which are the exponents for the short-range ($R=1$)
  case. }
\label{fig:simul}
\end{figure}

\section{Critical properties}
\label{app4}


\subsection{The critical threshold}
For a generic extended-range percolation process with range $R$, the size of the EGCC is given by
\begin{equation}
    S=\phi\left[1-g_0\left(\phi u_1+(1-\phi)u_2 \right) \right],
\end{equation}
where $u_1$ and $u_2$ are, in general, solutions of a nonlinear system of $R$ equations of the form
\begin{equation}
    \boldsymbol{u}=\boldsymbol{F}(\boldsymbol{u};\phi),
    \label{eq:fixed_point}
\end{equation}
with $\boldsymbol{u}$ an $R-$component vector $\boldsymbol{u}=(u_1,u_2,u_3,u_4, u_5, u_6,\dots)$ and
$\boldsymbol{F}(\boldsymbol{u};\phi)=(F_1(\boldsymbol{u};\phi),F_2(\boldsymbol{u};\phi),\dots,F_R(\boldsymbol{u};\phi))$.
Eq.~\eqref{eq:fixed_point} can be solved by iterating the recursive equation $\boldsymbol{u}(t+1)=\boldsymbol{F}(\boldsymbol{u}(t);\phi)$.
Note that $\boldsymbol{u}^{\#}(\phi)=(1,1,1,1,g_1(1-\phi), g_1(1-\phi), \dots)$ is always a solution for any value of $R$ and for any $\phi$.
This solution corresponds to $S=0$, the non-percolating phase.
The solution $\boldsymbol{u}^{\#}$ is stable until $\phi$ reaches a critical value
$\phi_c^{(R)}$: at this point, $\boldsymbol{u}^{\#}$ is marginally stable
and another fixed point $\boldsymbol{u^*}$ appears,
which is attractive for $\phi > \phi_c^{(R)}$.
To analyze the stability of the trivial fixed point, we linearize the equations by setting
$\boldsymbol{u}=\boldsymbol{u}^{\#}-\boldsymbol{\epsilon}$, from which it
follows, using the fact that
$\boldsymbol{u}^{\#}=\boldsymbol{F}(\boldsymbol{u}^{\#};\phi)$
\begin{align}
    \boldsymbol{\epsilon}(t+1)\simeq \hat{J}(\boldsymbol{u}^{\#};\phi)\boldsymbol{\epsilon}(t)
\end{align}
where
$\hat{J}(\boldsymbol{u}^{\#};\phi)=D\boldsymbol{F}(\boldsymbol{u}^{\#};\phi)$ is
the Jacobian matrix evaluated at the trivial fixed point.  It follows
that the trivial fixed point $\boldsymbol{u}^{\#}$, corresponding to
$\boldsymbol{\epsilon}=0$, is an attractive solution if
$\rho(\hat{J})<1$, where $\rho(\hat{J})$ is the spectral radius of
$\hat{J}$.  Furthermore, the Perron-Frobenius theorem \cite{horn2012matrix}
tells us that the $\rho(\hat{J})=\lambda_1$, where $\lambda_1$ is the largest
real eigenvalue of $\hat{J}$. Hence we can conclude that the critical threshold
$\phi_c^{(R)}$ is the value such that
\begin{equation}
    \det \left[\hat{J}(\boldsymbol{u}^{\#}, \phi_c^{(R)})-\mathbb{1} \right]= 0,
    \label{eq:critical_point}
\end{equation}
that is, when $\lambda_1$ equals $1$.
At this point, $\boldsymbol{u}^{\#}$ becomes unstable and the other
nontrivial fixed point $\boldsymbol{u^*}$, corresponding to a
percolating phase with $S>0$, appears. Eq.~\eqref{eq:critical_point}
allows us to find the critical point for any $R$.
Note that this argument holds if $g_1'(1)$ exists, which means that the
branching factor $b=\langle k(k-1)\rangle/\langle k \rangle$
must be finite.
If instead the degree distribution has $b=\infty$, since for $R=1$
the threshold vanishes, we can conclude that $\phi_c^{(R)}=0$ for any $R$.

\subsection{The exponent $\beta$}
In this subsection we derive the value of the critical exponent $\beta$,
determining the singularity of the EGCC size at the critical point
\begin{equation}
   S \sim (\phi-\phi^{(R)}_c)^{\beta}
\end{equation}
The procedure is as follows:
\begin{enumerate}
    \item we set $\phi=\phi^{(R)}_c+\delta \phi$, and consider small $\delta \phi$:
    \item we expand the equations for $u_i$ around the trivial fixed point $u_i=u^{\#}_i-\epsilon_i$;
      it is sufficient to expand up to the two lowest orders in $\epsilon_i$;
    \item we find the position of the nontrivial fixed point
      $\boldsymbol{\epsilon^*} \neq 0$ up to the lowest order in $\delta \phi$;
    \item we expand $S$ above $\phi^{(R)}_c$ for small $\epsilon_i$ (since
      $\delta \phi$ is small) and we finally get an expression of the
      form $S \sim \delta \phi^\beta$.
\end{enumerate}
Of particular interest is the case of power-law degree distributed networks with $p(k) \sim k^{-\gamma}\theta(k - k_{\text{min}})$, where $\theta(x)$ is the Heaviside step function, for which the procedure described above must be carried out carefully.
The main tool is the asymptotic expansion for the
generating functions close to their singular point $z=1$. The generating functions for power-law degree distribution $p(k)$ and excess degree distribution $q(r)=(r+1)p(r+1)/\langle k \rangle$, within the continuous-degree approximation, are
\begin{align}
\label{eq:gen_0}
  g_0(z) &= \int_{0}^{\infty}dk p(k)z^k =  (\gamma-1)k_{\text{min}}^{\gamma-1}\int_{k_{\text{min}}}^{\infty}dk k^{-\gamma}z^k = (\gamma-1)\frac{\Gamma(1-\gamma, k_{\text{min}}\psi)}{(k_{\text{min}}\psi)^{1-\gamma}},\\
  \label{eq:gen_1}
  g_1(z) &= \int_{0}^{\infty}dr q(r)z^r = (\gamma-2)k_{\text{min}}^{\gamma-2}\int_{k_{\text{min}}}^{\infty}dk k^{-(\gamma-1)}z^{k-1}= \frac{(\gamma-2)}{z}\frac{\Gamma(2-\gamma, k_{\text{min}}\psi)}{(k_{\text{min}}\psi)^{2-\gamma}},
\end{align}
where $\psi=\ln(1/z)$ and $\Gamma(a,x)$ is the Incomplete Gamma
function. For $z \simeq 1$ the functions have a singular behavior
which depends on the value of $\gamma$.
The asymptotic expansion which keeps only the lowest order terms is presented in Appendix \ref{app7}.

\subsection{The exponent $\beta$ for $R=2$}
Setting $u_1=1-\epsilon_1$ and $u_2=1-\epsilon_2$, we have
\begin{align}
    S&=\phi(1-g_0(\phi u_1+(1-\phi)u_2)=\phi[1-g_0(1-(\phi \epsilon_1 +(1-\phi)\epsilon_2)],
\end{align}
where $\epsilon_1$ and $\epsilon_2$ are the solutions of the equations
\begin{align}
    1-\epsilon_1&=g_1(\phi u_1 + (1-\phi) u_2)=g_1(1-(\phi \epsilon_1 +(1-\phi)\epsilon_2),\\
    1-\epsilon_2&=g_1(\phi u_1 +1-\phi)=g_1(1-\phi \epsilon_1).
\end{align}
Using Eq.~\eqref{eq:as_gen_0} and Eq.~\eqref{eq:as_gen_1} we get
\begin{align}
S\simeq \left< k \right> \phi[\phi\epsilon_1+(1-\phi)\epsilon_2]
\label{eq:S_2_range}
\end{align}
and
\begin{align}
    \epsilon_1 &\simeq B\phi \epsilon_1+B(1-\phi)\epsilon_2-\frac{D}{2}\left[ \phi^2\epsilon_1^2+2\phi(1-\phi)\epsilon_1 \epsilon_2 +(1-\phi)^2\epsilon_2^2\right]-C(\gamma-2)(\phi \epsilon_1+(1-\phi)\epsilon_2)^{\gamma-2},\\
    \epsilon_2 &\simeq B\phi \epsilon_1 -\frac{1}{2}D \phi^2 \epsilon_1^2-C(\gamma-2)\phi^{\gamma-2}\epsilon_1^{\gamma-2}.
\end{align}
See Appendix \ref{app7} for the values of $\langle k \rangle$, $B$, $D$ and $C(\gamma-2)$.
The solution of these equations for $\phi \simeq \phi_c^{(R=2)}$ depends on the value of $\gamma$.

\subsubsection{Non-heterogeneous networks, i.e., $\gamma>4$ or ER networks}
The leading terms in the equation for $\epsilon_2$ are those of order
$\epsilon_1$ and $\epsilon_1^2$. Keeping only these two lowest
order terms and inserting them into the equation for $\epsilon_1$, we get
\begin{equation}
\Lambda(\phi) \epsilon_1 + \Omega \epsilon_1^2=0
\end{equation}
where we set
\begin{align}
    \Lambda(\phi)&=-(B\phi-1)-B^2\phi(1-\phi),\\
    \Omega&=\frac{D \phi^2}{2}\left[(B\large(1-\phi)+1\large)^2+B(1-\phi) \right].
\end{align}
which admits a nontrivial solution $\epsilon_1^*>0$
\begin{equation}
\epsilon_1^* = \left[\frac{-\Lambda(\phi)}{\Omega} \right]\simeq-\left[ \frac{\Lambda(\phi_c)+\Lambda'(\phi_c)\delta \phi}{\Omega}\right] =\left[ -\frac{\Lambda'(\phi_c)\delta \phi}{\Omega}\right]\sim {\delta \phi}
\end{equation}
as soon as $\phi>\phi_c$. We can conclude that both $\epsilon_1$ and
$\epsilon_2$ are of $\mathcal{O}(\delta \phi)$, hence we get from
Eq.~\eqref{eq:S_2_range} $\beta=1$. Notice that the same result holds for any uncorrelated random graph with finite $\langle k^3 \rangle$, such as ER networks.
This result is confirmed in Fig. \ref{fig:simul}(c).

\subsubsection{Weakly-heterogeneous networks, i.e., $3<\gamma<4$}
In this case the leading terms in the equation for $\epsilon_2$ are
those of order $\epsilon_1$ and $\epsilon_1^{\gamma-2}$. Substituting $\epsilon_2$
in the equation for $\epsilon_1$ and keeping only lowest order
terms gives
\begin{equation}
    \Lambda(\phi)\epsilon_1+E \epsilon_1^{\gamma-2}=0,
\end{equation}
where $\Lambda(\phi)$ is defined as before and
\begin{equation}
E=C(\gamma-2)\phi^{\gamma-2}\left[\large(1+B(1-\phi) \large)^{\gamma-2}+ B(1-\phi) \right].
\end{equation}
Hence the nontrivial solution $\epsilon_1^*$ is
\begin{equation}
\epsilon_1^* = \left[\frac{-\Lambda(\phi)}{E}\right]^{1/(\gamma-3)}\simeq\left[-\frac{\Lambda(\phi_c)+\Lambda'(\phi_c)\delta \phi}{E}\right]^{1/(\gamma-3)}\sim (\delta \phi)^{1/(\gamma-3)}.
\end{equation}
We conclude that both $\epsilon_1$ and $\epsilon_2$ are $\mathcal{O}((\delta \phi) ^{1/(\gamma-3)})$.
From Eq.~\eqref{eq:S_2_range} it follows that $\beta=1/(\gamma-3)$.
This result is confirmed in Fig. \ref{fig:simul}(d).

\subsubsection{Strongly-heterogenous networks: $2<\gamma<3$}

In this case $\delta \phi=\phi$, since $\phi_c=0$.
The leading terms in the equation for $\epsilon_2$ are again the ones
considered in the case $3<\gamma<4$, but now the leading order is
$(\phi\epsilon_1)^{\gamma-2}$. This implies that, in the equation for $\epsilon_1$,
the last term gives at leading order a contribution
\begin{align}
\nonumber
\left[\phi\epsilon_1+(1-\phi)\epsilon_2\right]^{\gamma-2}&\simeq \left[\phi\epsilon_1+(1-\phi)\large(B\phi\epsilon_1-C(\gamma-2)\phi^{\gamma-2}\epsilon_1^{\gamma-2} \large)\right]^{\gamma-2}\\ \nonumber
&=\left[\phi\epsilon_1+B\phi(1-\phi)\epsilon_1-(1-\phi)C(\gamma-2)\phi^{\gamma-2}\epsilon_1^{\gamma-2} \right]^{\gamma-2}\\ \nonumber
&=\left[-(1-\phi)C(\gamma-2)\phi^{\gamma-2}\epsilon_1^{\gamma-2} \right]^{\gamma-2}\left[1-\frac{1+B(1-\phi)}{(1-\phi)C(\gamma-2)\phi^{\gamma-2}}(\phi\epsilon_1)^{3-\gamma} \right]^{\gamma-2} \\ \nonumber
&\simeq\left[-(1-\phi)C(\gamma-2)\phi^{\gamma-2}\epsilon_1^{\gamma-2} \right]^{\gamma-2}\left[1-({\gamma-2})\frac{1+B(1-\phi)^2}{(1-\phi)C(\gamma-2)\phi^{\gamma-2}}(\phi\epsilon_1)^{3-\gamma}+\dots \right]\\ \nonumber
&\simeq\left[-(1-\phi)C(\gamma-2)\phi^{\gamma-2}\epsilon_1^{\gamma-2} \right]^{\gamma-2}\\
&=\left[-(1-\phi)C(\gamma-2)\right]^{\gamma-2}\phi^{(\gamma-2)^2}\epsilon_1^{(\gamma-2)^2}.
\end{align}
Furthermore, since $\gamma-2<1$, the terms of order
$\epsilon_2^{(\gamma-2)^2}$ dominate with respect to those
of order $\epsilon_1^{\gamma-2}$ appearing in the equation
for $\epsilon_1$.
With this in mind, we get from the equation for $\epsilon_1$
\begin{equation}
\epsilon_1 \simeq B \phi \epsilon_1+B^2\phi(1-\phi)\epsilon_1-B(1-\phi)C(\gamma-2)\phi^{\gamma-2}\epsilon_1^{\gamma-2}-C(\gamma-2)\left[-(1-\phi)C(\gamma-2)\phi^{\gamma-2}\epsilon_1^{\gamma-2} \right]^{\gamma-2}
\end{equation}
which can be rewritten as
\begin{equation}
\epsilon_1 \simeq \frac{B(1-\phi)C(\gamma-2)}{\Lambda(\phi)}\phi^{\gamma-2}\epsilon_1^{\gamma-2}+\left[\frac{-C(\gamma-2)\Large(-(1-\phi)C(\gamma-2)\Large)^{\gamma-2}}{\Lambda(\phi)}\right]\phi^{(\gamma-2)^2}\epsilon_1^{(\gamma-2)^2},
\end{equation}
In the last expression, the exponent $(\gamma-2)^2$ is smaller than
$\gamma-2$. Hence only the last term on the right hand side must be kept
\begin{equation}
\epsilon_1  \simeq A \epsilon_1^{\alpha},
\end{equation}
where we set
\begin{align}
\alpha &= (\gamma-2)^2,\\
A &= \left[\frac{-C(\gamma-2)\Large(-(1-\phi)C(\gamma-2)\Large)^{\gamma-2}}{\Lambda(\phi)}\right]\phi^{\alpha}.
\end{align}
As a consequence
\begin{equation}
\epsilon_1^* = A^{1/(1-\alpha)}\sim \phi^{(\gamma-2)^2/[1-(\gamma-2)^2]}.
\end{equation}
From the equation for $\epsilon_2$ we get
\begin{equation}
    \epsilon_2^* \sim \phi^{\gamma-2}\left(\phi^{(\gamma-2)^2/[1-(\gamma-2)^2]}\right)^{\gamma-2}\sim \phi^{(\gamma-2)/[1-(\gamma-2)^2]}.
\end{equation}
From Eq.~\eqref{eq:S_2_range} we have
\begin{align}
\nonumber
    S &\simeq \left< k \right>\phi^2 \epsilon_1 + \left< k \right>\phi \epsilon_2 \sim \phi^{2+(\gamma-2)^2/[1-(\gamma-2)^2]}+\phi^{1+(\gamma-2)/[1-(\gamma-2)^2]}\\
    &\sim \phi^{1+(\gamma-2)/[1-(\gamma-2)^2]},
\end{align}
which implies
\begin{equation}
    \beta_{(R=2)} = 1+\frac{(\gamma-2)}{[1-(\gamma-2)^2]}.
\end{equation}
Remarkably, while for $\gamma>3$ we recover the same exponent of standard percolation,
for $2<\gamma<3$ we find a new nontrivial dependence of $\beta$ on $\gamma$, different from the one valid
for $R=1$.

\subsection{The exponent $\beta$ for $R>2$}
For $\gamma>3$ the exponent $\beta$ is the same as for standard $R=1$ percolation even for $R>2$.
This can be physically understood by considering that for $\gamma>3$
the extended-neighborhood always involves a large but finite number of nodes.
Hence the $R$-range process cannot differ in the universal properties, i.e. in the
critical exponents, from the $R=1$ case. From a mathematical point of
view, one can easily check that all $\epsilon_i$ are of the same order,
and hence the same scaling of $\epsilon_1$ versus $\delta \phi$ holds.
As a consequence $S$ scales in the same way for any $R$
(see Fig.~\ref{fig:simul} (c) and (d)).

In the case $2<\gamma<3$, instead, the extended-neighborhood of a node
reaches a diverging number of other nodes and this qualitatively changes
the properties of the process.

Let us consider $R=6$.
Noticing that $\phi u_1 \ll u_3 $ for $\phi \sim 0$, it is easy to see that the equation for $u_6=g_1(1-\phi)-\epsilon_6$, Eq.~\eqref{eq:w_5}, yields, at lowest order
\begin{equation}
\epsilon_6 \simeq g_1(1-\phi)-g_1\big(g_1(1-\phi)\big) \simeq -C(\gamma-2)\big[-C(\gamma-2)\phi^{\gamma-2}\big]^{\gamma-2} \sim \phi^{(\gamma-2)^2}.
\end{equation}
From Eq.~\eqref{eq:w_4_} for $u_5=g_1(1-\phi)-\epsilon_5$, we get
\begin{equation}
\epsilon_5 \simeq g_1(1-\phi)-g_1\big((1-\phi)u_6\big) \sim \phi^{(\gamma-2)^3}.
\end{equation}
Considering Eq.~\eqref{eq:u4} for $u_4=1-\epsilon_4$ it is also easy to realize that
the leading contribution is provided by the first term on the r.h.s.,
$u_5$, which, expanded for small $\phi$ gives
\begin{equation}
\epsilon_4 \sim \epsilon_5 \sim \phi^{(\gamma-2)^3}.
\end{equation}
In Eq.~\eqref{eq:u_3} for $u_3$
the leading contribution on the r.h.s. in the limit of small $\phi$,
is the first, thus giving
\begin{equation}
\epsilon_3 \sim \epsilon_4^{\gamma-2} \sim \phi^{(\gamma-2)^4}.
\end{equation}
By the same token, inspection of the equations for $u_2$ and $u_1$
promptly leads to the conclusion that $\epsilon_1 \sim \epsilon_2^{\gamma-2} \sim \epsilon_3^{(\gamma-2)^2}$.

From Eq.~\eqref{eq:S_2_range} we get
\begin{equation}
  S \sim \phi \epsilon_2 \sim \phi^{1+(\gamma-2)^5},
\end{equation}
implying $\beta_{(R=6)}=1+(\gamma-2)^5$.
Using the same type of argument, it is easy to recognize for $R=5$ that $\epsilon_4 \sim \epsilon_5 \sim \phi^{(\gamma-2)^2}$ and $\epsilon_i \sim \epsilon_{i+1}^{\gamma-2}$ for $i < 4$.  Similarly, we find that for $R=3$ and $R=4$ the relations $\epsilon_R \sim \phi^{(\gamma-2)}$ and
$\epsilon_i \sim \epsilon_{i+1}^{\gamma-2}$ for $1 \leq i<R$ hold. Summing up we find, for $R=3,4,5,6$,
\begin{equation}
\beta_{(R)} = 1+(\gamma-2)^{R-1}.
\label{eq:conj_beta}
\end{equation}

These values of $\beta$ obey the expectation that the larger $R$, the
closest $\beta$ is to $1$, which is the pure mean-field value.

We would expect Eq.~\eqref{eq:conj_beta} to hold also for $R>6$. This conjecture can be physically justified by noticing that for small values of $\phi$, the largest contribution to the probabilities of not reaching the EGCC always comes from configurations in which all the outgoing neighbors are inactive, since it is very unlikely to encounter an active node. Hence, even if we don't know the recursive equations for the probabilities $u_i$ for $i>6$, we can argue that scaling relations between $\epsilon_i$ and $\epsilon_{i+1}$, similar to the ones we found for $R \leq 6$, hold, leading to Eq.~\eqref{eq:conj_beta}.

\section{Message-passing equations for $R\leq4$}
\label{app5}

Based on Eqs. (1), (2), (4), (5) and (6) of the main text, it is straightforward to write the
corresponding message-passing equations for $R=4$,
\begin{align}
u^{(i \leftarrow j)}_1 &= \prod_{k \in \partial_j \setminus i} \left[ \phi u^{(j \leftarrow k)}_1 + (1-\phi) u^{(j \leftarrow k)}_2  \right],  \label{eq2.10}  \\
u^{(i \leftarrow j)}_2 &= \prod_{k \in \partial_j \setminus i} \left[ \phi u^{(j \leftarrow k)}_1 + (1-\phi) u^{(j \leftarrow k)}_3  \right],  \label{eq2.20}  \\
u^{(i \leftarrow j)}_3 &= \prod_{k \in \partial_j \setminus i} \left[ \phi u^{(j \leftarrow k)}_1 + (1-\phi) u^{(j \leftarrow k)}_3  \right] + \prod_{k \in \partial_j \setminus i} \left[ (1-\phi) u^{(j \leftarrow k)}_4  \right] - \prod_{k \in \partial_j \setminus i} \left[ (1-\phi) u^{(j \leftarrow k)}_3  \right],  \label{eq2.30}  \\
u^{(i \leftarrow j)}_4 &= \prod_{k \in \partial_j \setminus i} \left[ \phi u^{(j \leftarrow k)}_1 + (1-\phi) u^{(j \leftarrow k)}_3  \right] + \prod_{k \in \partial_j \setminus i} \left( 1-\phi  \right) - \prod_{k \in \partial_j \setminus i} \left[ (1-\phi) u^{(j \leftarrow k)}_3  \right].  \label{eq2.40}
\end{align}

\noindent
The expression for the relative size of the EGCC remains the same as for $R=2$,
\begin{align}
S = \frac{1}{N} \sum_{i=1}^N S_i,  \label{eq2.44}
\end{align}
\noindent
with
\begin{align}
S_i = \phi \left[ 1- \prod_{j \in \partial_i} \left( \phi u^{(i \leftarrow j)}_1 + (1-\phi) u^{(i \leftarrow j)}_2  \right) \right].  \label{eq2.45}
\end{align}
\noindent
The critical threshold, $\phi_c^{(R=4)}$, may be found using the same procedure outlined in the main text. We must linearize Eqs. (\ref{eq2.10}), (\ref{eq2.20}), (\ref{eq2.30}) and (\ref{eq2.40}) around the trivial solution $u^{(i \leftarrow j)}_1 = u^{(i \leftarrow j)}_2 = u^{(i \leftarrow j)}_3 = u^{(i \leftarrow j)}_4 = 1$, to obtain
\begin{align}
\epsilon^{(i \leftarrow j)}_1 &= \phi \sum_{k \in \partial_j \setminus i} \epsilon^{(j \leftarrow k)}_1 + (1-\phi) \sum_{k \in \partial_j \setminus i} \epsilon^{(j \leftarrow k)}_2,  \label{eq2.50}  \\
\epsilon^{(i \leftarrow j)}_2 &= \phi \sum_{k \in \partial_j \setminus i} \epsilon^{(j \leftarrow k)}_1 + (1-\phi) \sum_{k \in \partial_j \setminus i} \epsilon^{(j \leftarrow k)}_3,  \label{eq2.60}  \\
\epsilon^{(i \leftarrow j)}_3 &= \phi \sum_{k \in \partial_j \setminus i} \epsilon^{(j \leftarrow k)}_1 + (1-\phi) \sum_{k \in \partial_j \setminus i} \epsilon^{(j \leftarrow k)}_3 + (1-\phi)^{q_j-1} \sum_{k \in \partial_j \setminus i} \epsilon^{(j \leftarrow k)}_4 - (1-\phi)^{q_j-1} \sum_{k \in \partial_j \setminus i} \epsilon^{(j \leftarrow k)}_3,  \label{eq2.70}  \\
\epsilon^{(i \leftarrow j)}_4 &= \phi \sum_{k \in \partial_j \setminus i} \epsilon^{(j \leftarrow k)}_1 + (1-\phi) \sum_{k \in \partial_j \setminus i} \epsilon^{(j \leftarrow k)}_3 - (1-\phi)^{q_j-1} \sum_{k \in \partial_j \setminus i} \epsilon^{(j \leftarrow k)}_3,  \label{eq2.80}
\end{align}

\noindent
where $q_j$ denotes the degree of node $j$, and $\epsilon^{(i \leftarrow j)}_r = 1 - u^{(i \leftarrow j)}_r \ll 1$ for $r\leq4$. The Jacobian matrix $\hat{C}$ obtained from these equations can be written as
\begin{align}
\hat{C} =
\begingroup 
\setlength\arraycolsep{8pt}
\begin{pmatrix}
 \phi \hat{B} & (1-\phi) \hat{B} & 0 & 0 \\
 \phi \hat{B} & 0 & (1-\phi) \hat{B} & 0 \\
 \phi \hat{B} & 0 & (1-\phi) \hat{B} - \hat{M}^{(\phi)} & \hat{M}^{(\phi)} \\
 \phi \hat{B} & 0 & (1-\phi) \hat{B} - \hat{M}^{(\phi)} & 0
\end{pmatrix}
\endgroup,
\label{eq2.100}
\end{align}

\noindent
which is a nonnegative matrix. [The matrix $\hat{M}^{(\phi)}$ is defined as $M^{(\phi)}_{i \leftarrow j, k \leftarrow l} = (1-\phi)^{q_j-1} \delta_{jk} (1-\delta_{il})$.] Using the Perron-Frobenius theorem, the largest real eigenvalue of matrix $\hat{C}$ must be $1$ at the critical threshold $\phi = \phi_c^{(R=4)}$, which can be found by numerical means. As explained in the main text, for $R<4$ the same procedure may be applied, only one must consider the appropriate $R \times R$ matrix in the top left corner of matrix $\hat{C}$.

\section{Real-world networks}
\label{app6}

In Fig. 2(d) of the main text we compare the solutions of the message-passing equations for $R=2$ in three real-world networks with simulations results. Table I presents the number of nodes $N$, the mean degree $\langle k \rangle$, the mean clustering coefficient $C$ and the critical threshold $\phi_c^{(R=2)}$ (as determined by the message-passing equations), as well as the source for each network.

\begin{table}[h!]
\centering
\begin{tabular}{ |l|c|c|c|c|c|c|c| }
 \hline
 \textbf{Network} & Description & $N$ & $\langle k \rangle$ & C & $\phi_c^{(R=2)}$ \\
 \hline
 Internet \footnotemark[1] & Snapshot of the structure of the Internet at the level of autonomous systems & $22963$ & $4.22$ & $0.2304$ & $2.41 \times 10^{-4}$ \\
 \hline
 Social \footnotemark[1] & Condensed matter collaboration network in the period 1995-2003 & $27519$ & $8.44$ & $0.6461$ & $6.57 \times 10^{-4}$ \\
 \hline
 P2P \footnotemark[2] & Gnutella peer-to-peer network on Aug. 31, 2002 & $62561$ & $4.73$ & $0.0055$ & $7.04 \times 10^{-3}$ \\
 \hline
\end{tabular}
\footnotetext[1]{Downloaded from: \url{http://www-personal.umich.edu/~mejn/netdata/}.}
\footnotetext[2]{Downloaded from: \url{http://snap.stanford.edu/data/}.}


\caption{Number of nodes $N$, mean degree $\langle k \rangle$, mean clustering coefficient $C$ and the $2$-range percolation threshold $\phi_c^{(R=2)}$ of the real-world networks considered in the main text.}
\label{table1}
\end{table}


%


\section{Asymptotic expansion for the generating functions}
\label{app7}

The generating functions for power-law degree distributions (within the continuous-degree approximation) can be expressed in terms of the Incomplete Gamma Function as in Eqs.\eqref{eq:gen_0} and \eqref{eq:gen_1}. For $\gamma<4$, we cannot naively Taylor expand the generating
functions up to quadratic order, because of the singularity in $z=1$,
which is exactly the point in which we need to evaluate the generating
function and its derivatives. However, we can use the asymptotic
expansion for the Incomplete Gamma Function \cite{bender1999advanced}
\begin{equation}
\frac{\Gamma(a,x)}{x^a}=\frac{\Gamma(a)}{x^a}-\sum_{k=0}^{\infty}\frac{(-1)^kx^k}{k!(a+k)}
\end{equation}
for $x \sim 0$.  Setting $z=1-\varepsilon$, substituting $1/z \simeq
1+\varepsilon+\varepsilon^2$ and $\psi \simeq \varepsilon +
\varepsilon^2/2$, using the expansion above with
$x=k_{\text{min}}(\varepsilon + \varepsilon^2/2)$, we get from
Eq.~\eqref{eq:gen_0}
\begin{align}
\nonumber
g_0(1-\varepsilon) &\simeq (\gamma-1)\left[\frac{\Gamma(1-\gamma)}{[k_{\text{min}}(\varepsilon+\varepsilon^2/2)]^{1-\gamma}}-\frac{1}{1-\gamma}+\frac{k_{\text{min}}(\varepsilon+\varepsilon^2/2)}{2-\gamma}-\frac{k_{\text{min}}^2(\varepsilon+\varepsilon^2/2)^2}{2(3-\gamma)} \right] \\ \nonumber
& \simeq (\gamma-1)\left[\Gamma(1-\gamma)(k_{\text{min}}\varepsilon)^{\gamma-1}-\frac{1}{1-\gamma}+\frac{k_{\text{min}}(\varepsilon+\varepsilon^2/2)}{2-\gamma}-\frac{k_{\text{min}}^2\varepsilon^2}{2(3-\gamma)} \right] \\ \nonumber
&=1+C(\gamma-1)\varepsilon^{\gamma-1}-\frac{\gamma-1}{\gamma-2}k_{\text{min}}\varepsilon+\frac{1}{2}\left[\frac{\gamma-1}{\gamma-3}k_{\text{min}}^2-\frac{\gamma-1}{\gamma-2}k_{\text{min}} \right]\varepsilon^2,
\end{align}
where $C(a)=a\Gamma(-a)k_{\text{min}}^a$.
Following the same computations, from Eq.~\eqref{eq:gen_1} it follows
\begin{align}
\nonumber
    g_1(1-\varepsilon)&\simeq (\gamma-2)(1+\varepsilon+\varepsilon^2 )\left[\frac{\Gamma(2-\gamma)}{k_{\text{min}}^{2-\gamma}(\varepsilon +\varepsilon^2/2)^{2-\gamma}}-\frac{1}{2-\gamma}+\frac{k_{\text{min}}(\varepsilon +\varepsilon^2/2)}{3-\gamma}-\frac{k_{\text{min}}^{2}(\varepsilon +\varepsilon^2/2)^2}{2(4-\gamma)} \right]
    \\ \nonumber
    &= (\gamma-2)(1+\varepsilon+\varepsilon^2 )\left[{\Gamma(2-\gamma)}k_{\text{min}}^{\gamma-2}\varepsilon^{\gamma-2}{(1+\varepsilon/2)^{\gamma-2}}-\frac{1}{2-\gamma}+\frac{k_{\text{min}}(\varepsilon +\varepsilon^2/2)}{3-\gamma}-\frac{k_{\text{min}}^2(\varepsilon +\varepsilon/2)^2}{2(4-\gamma)} \right]\\ \nonumber
    & \simeq (\gamma-2)(1+\varepsilon+\varepsilon^2 )\left[{\Gamma(2-\gamma)}(k_{\text{min}}\varepsilon)^{\gamma-2}-\frac{1}{2-\gamma}+\frac{k_{\text{min}}(\varepsilon +\varepsilon^2/2)}{3-\gamma}-\frac{k_{\text{min}}^2\varepsilon^2}{2(4-\gamma)} \right]\\ \nonumber
    &\simeq C(\gamma-2)\varepsilon^{\gamma-2}+1-\frac{\gamma-2}{\gamma-3}k_{\text{min}}(\varepsilon +\varepsilon^2/2)+\frac{\gamma-2}{2(\gamma-4)}k_{\text{min}}^2\varepsilon^2
    +\varepsilon -\frac{\gamma-2}{\gamma-3}k_{\text{min}}\varepsilon^2  +{\varepsilon^2} \\ \nonumber
    &=1-\left[\frac{\gamma-2}{\gamma-3}k_{\text{min}}-1 \right]\varepsilon+\frac{1}{2}\left[\frac{\gamma-2}{\gamma-4}k_{\text{min}}^2-3\frac{\gamma-2} {\gamma-3}k_{\text{min}}+2\right]\varepsilon^2 +C(\gamma-2)\varepsilon^{\gamma-2}. \nonumber
\end{align}

Summing up, we obtained the following expansions
\begin{align}
g_0(1-\varepsilon)&\simeq 1-\left<k \right>\varepsilon+\frac{1}{2}\left<k\right> B\varepsilon^2+C(\gamma-1)\varepsilon^{\gamma-1}, \label{eq:as_gen_0} \\
g_1(1-\varepsilon)&\simeq 1-B\varepsilon+\frac{1}{2}D\varepsilon^2+C(\gamma-2)\varepsilon^{\gamma-2}, \label{eq:as_gen_1}
\end{align}
where we defined
\begin{align}
C(a)&=a\Gamma(-a)k_{\text{min}}^a\\
\left<k \right> &= \frac{\gamma-1}{\gamma-2}k_{\text{min}},\\
B &= \frac{\gamma-2}{\gamma-3}k_{\text{min}}-1,\\
D &= \frac{\gamma-2}{\gamma-4}k_{\text{min}}^2-3\frac{\gamma-2} {\gamma-3}k_{\text{min}}+2.
\end{align}
Note that $B$ and $D$ correspond to ``true averages'', i.e., $B=\langle k(k-1)\rangle/\langle k \rangle$ and
$D=\langle k(k-1)(k-2)\rangle/\langle k \rangle$, respectively, only if the value of $\gamma$ is compatible
with the requirement for the average to be finite. In particular, $B$ is finite only if $\gamma>3$, and corresponds to the branching factor $b$. Notice that $C(\gamma-2)<0$ for $2<\gamma<3$, since $\Gamma(2-\gamma)<0$ in this range.

\end{widetext}


\bibliography{ext_perc_bib}

\end{document}